\documentclass[prb,twocolumn,showpacs,floatfix]{revtex4}
\usepackage{graphicx}% Include figure files
\usepackage{dcolumn}% Align table columns on decimal point
\usepackage{bm}% bold math
\usepackage{color}
\usepackage{ulem}
\begin{document}
\title{Origin of anomalous breakdown of Bloch's rule in the Mott-Hubbard insulator MnTe$_2$}
\author{Tapan Chatterji$^1$, Antonio M. dos Santos$^2$, Jamie J. Molaison$^2$, Thomas C. Hansen$^1$, Stefan Klotz$^3$ and Mathew Tucker$^4$}
\affiliation{$^1$Institut Laue-Langevin, B.P. 156, 38042 Grenoble Cedex 9, France\\
$^2$ Quantum Condensed Matter Div., Oak Ridge National Laboratory, TN 37831-6460, USA\\
$^3$IMPMC, CNRS UMR 7590, Universit\'e Pierre et Marie Curie, 75252 Paris\\
$^4$ISIS Facility, Rutherford Appleton Laboratory, Chilton, Didcot, UK}
\author{Kartik Samanta$^{5}$ and Tanusri Saha-Dasgupta$^{5}$}
\affiliation{$^5$ Department of Condensed Matter Physics and Materials Science, 
S.N. Bose National Centre for Basic Sciences, Kolkata-700098, India.}
\date{\today}
\begin{abstract}
We reinvestigate the pressure dependence of the crystal structure and antiferromagnetic phase transition in MnTe$_2$
by the rigorous and reliable tool of high pressure neutron powder diffraction. First-principles density functional theory calculations 
are carried out in order to gain microscopic insight. The measured N\'eel temperature of MnTe$_2$ is found to show unusually large 
pressure dependence of $12$ K GPa$^{-1}$. This gives rise to large violation of Bloch's rule given by $\alpha=\frac{d\log T_N}{d\log V}=-\frac{10}{3} \approx -3.3$, to a $\alpha$ value of -6.0 $\pm$ 0.1 for MnTe$_2$. The ab-initio calculation of the electronic
structure and the magnetic exchange interactions in MnTe$_2$, for the measured crystal structures at different pressures,
gives the pressure dependence of the Ne\'el temperature, $\alpha$ to be -5.61, in close agreement with experimental
finding. The microscopic origin of this behavior turns to be dictated by the distance dependence of the cation-anion
hopping interaction strength.
\end{abstract}
\maketitle

\section{Introduction}

Long time ago in 1966, Bloch \cite{bloch66} studied the pressure variation of the N\'eel temperature, $T_N$ and that of volume
($V$) of several transition-metal (TM) based antiferromagnetic insulators (AFI) and came up with the general
relationship
\begin{equation}
\alpha=\frac{d\log T_N}{d\log V}=-\frac{10}{3} \approx -3.3.
\end{equation}
In the localized-electron limit where perturbative superexchange theory is applicable, the N\'eel temperature can be
related to the effective TM-TM hopping interaction ($b$), charge transfer energy ($\Delta$) and Coulomb
interaction ($U$), as
\begin{equation}
T_N \sim b^2 \left[\frac{1}{U}+\frac{1}{2\Delta}\right].
\end{equation}
The first term in the above equation is the Anderson superexchange term and the second term involves the two electron
transfer from the anion.  A theoretical rationalization of the Bloch's rule comes from the calculations of the
variation of the cation-anion transfer integral $b^{ca}$ with the cation-anion bond length $r$, which varies as $r^{-n}$.
The calculated values\cite{smith69,shrivastava76} of $n$ using molecular orbital theory or configuration interaction method
on transition-metal oxides and flourides, turn out to be in the range 2.5-3. This leads to $T_N \sim r^{-10} \sim V^{-3.3}$,
assuming $b = (b^{ca})^{2}/\Delta$. 
Experimentally Bloch's rule is obeyed by a variety of Mott insulators. However, there are exceptions too. For example,
while Bloch rule is found to be obeyed in YCrO$_3$ and CaMnO$_3$, it was found to fail in LaMnO$_3$.\cite{goodenough}
The failure has been explained in terms of breakdown of localized approach used in Bloch's formulation. We note that the
cases discussed so far on the pressure dependence of T$_N$, all involve oxygen or flourine, {\it i.e.} anions
with 2$p$ electrons. As is well known, the nature of anionic wavefunction changes as one moves down the
column of the periodic table, from 2$p$ to 3$p$ series and even more to 4$p$ and 5$p$ series, effecting the TM-anion
bonding. It would therefore be of interest to consider the validity of Bloch's criterion in case of TM compounds
containing anions like Te.
\begin{figure}
\rotatebox{-90}{\resizebox{0.35\textwidth}{!}{\includegraphics{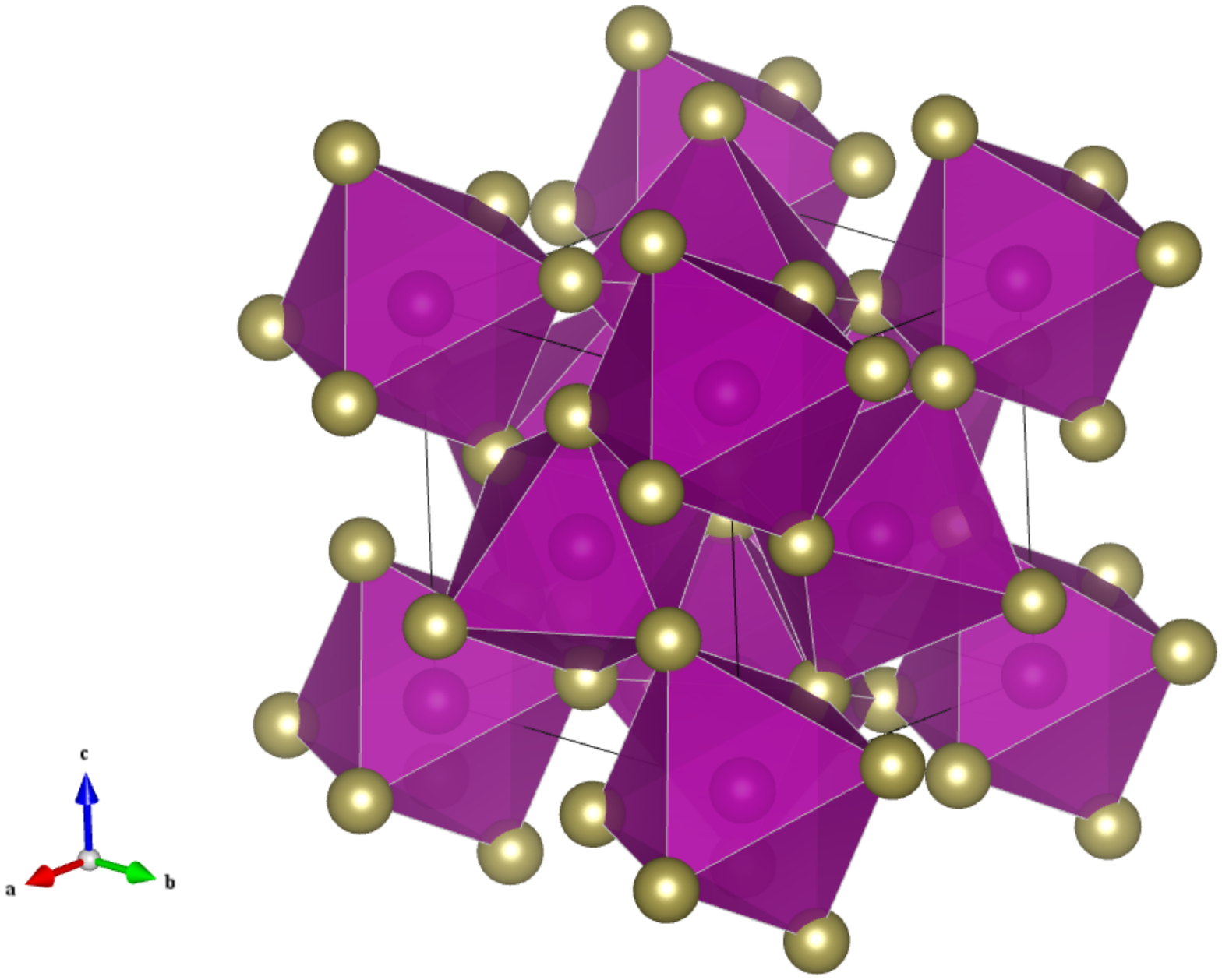}}}
\rotatebox{-90}{\resizebox{0.35\textwidth}{!}{\includegraphics{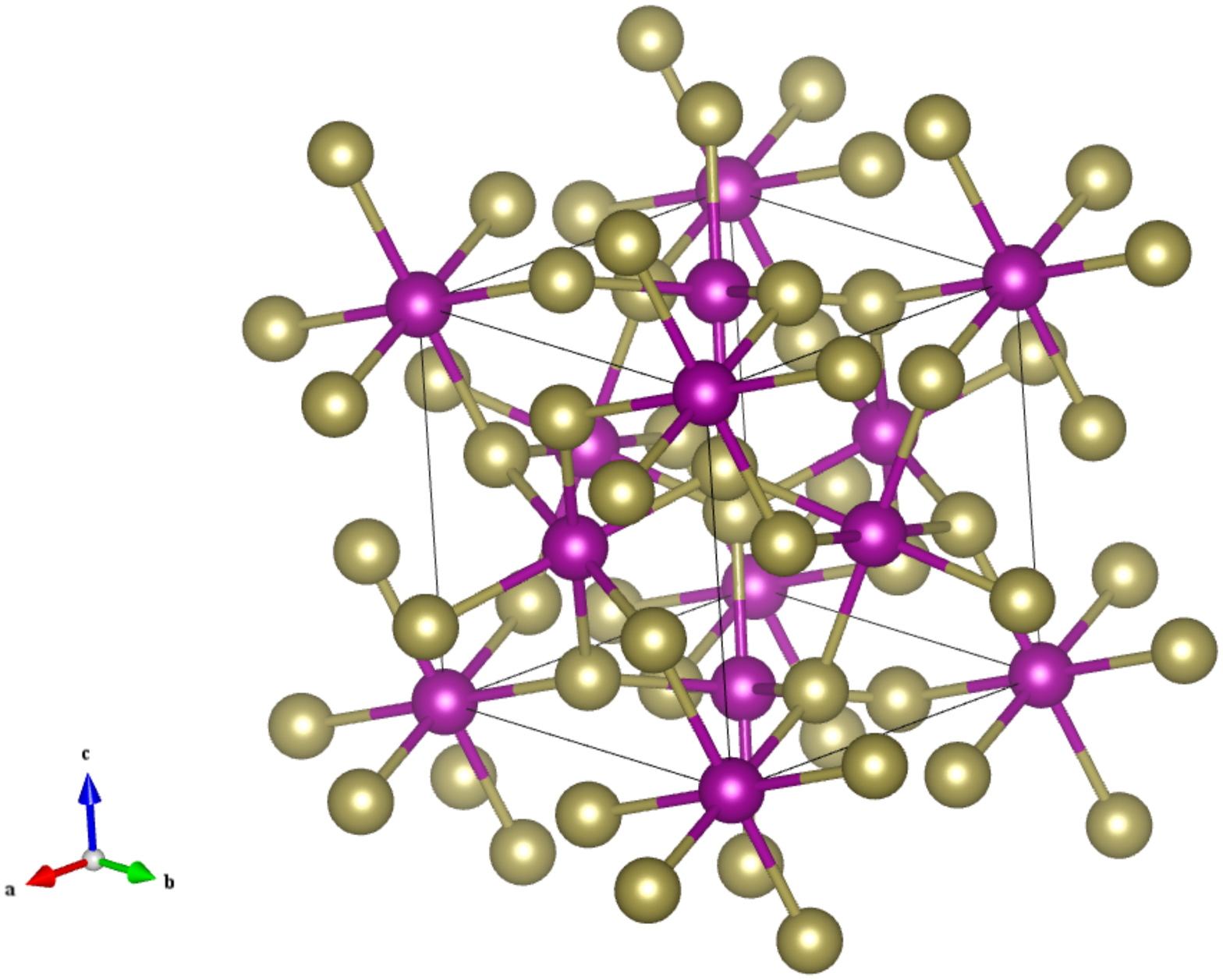}}}
\vspace{5pt}
 \caption {(Color online) Schematic representation of the crystal structure of MnTe$_2$. Mn atoms are shown as smaller purple spheres and Te atoms as larger yellow spheres. The upper panel shows the polyhedral representation of the MnTe$_6$ octahedra. The lower panel shows the ball and stick version of the same structure where the distorted octahedral coordination of the Mn by six Te atoms and distorted tetrahedral coordination of the Te atom by the three Mn and one Te atoms can be seen. }
 \label{structure}
\end{figure}

\begin{figure}
\resizebox{0.5\textwidth}{!}{\includegraphics{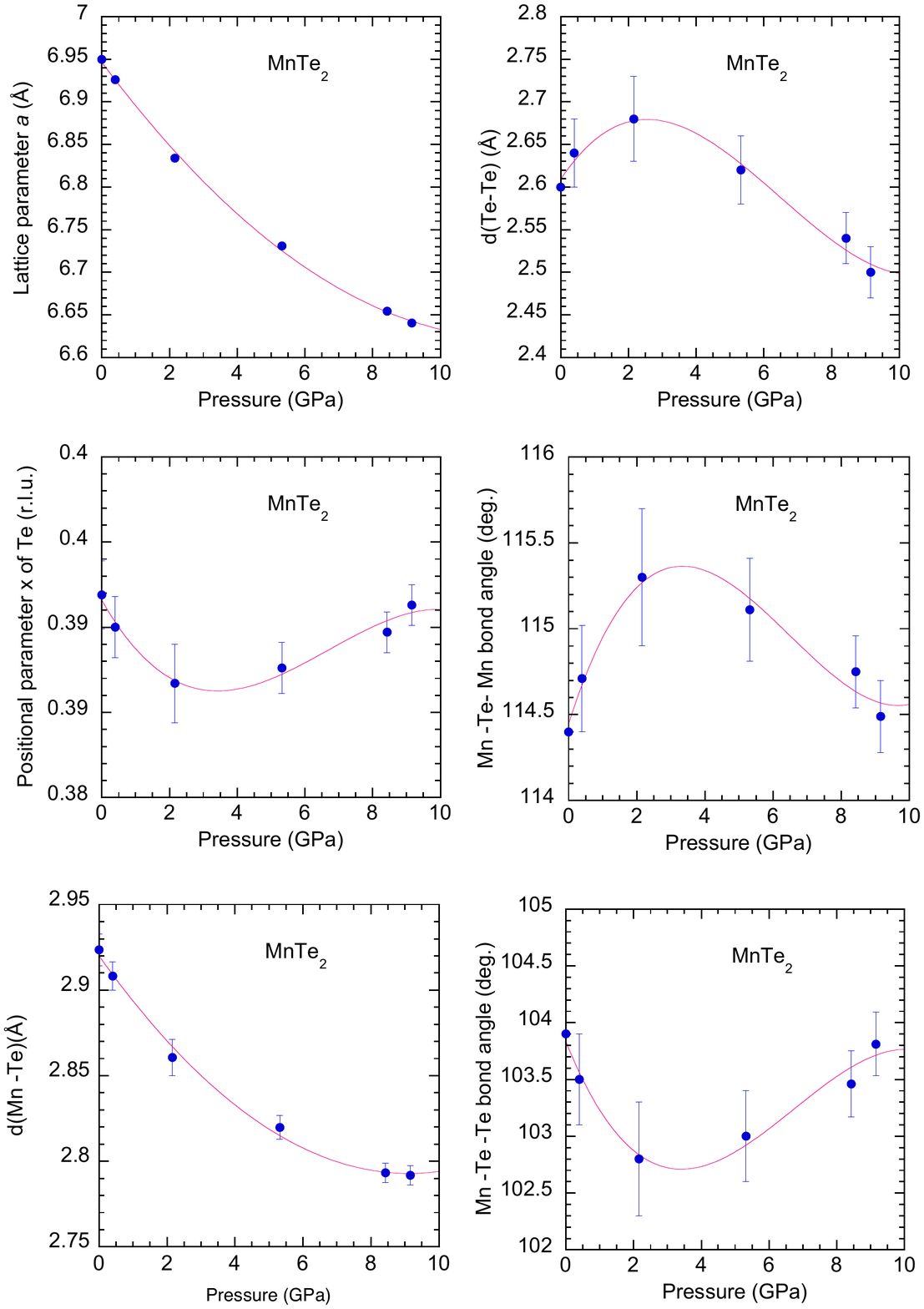}}
\vspace{5pt}
 \caption {(Color online) Pressure dependence of the structural parameters of MnTe$_2$ determined from the data measured on PEARL diffractometer of ISIS Facility. The red continuous lines are guides to the eye.}
 \label{param}
\end{figure}

\begin{figure}
\resizebox{0.5\textwidth}{!}{\includegraphics{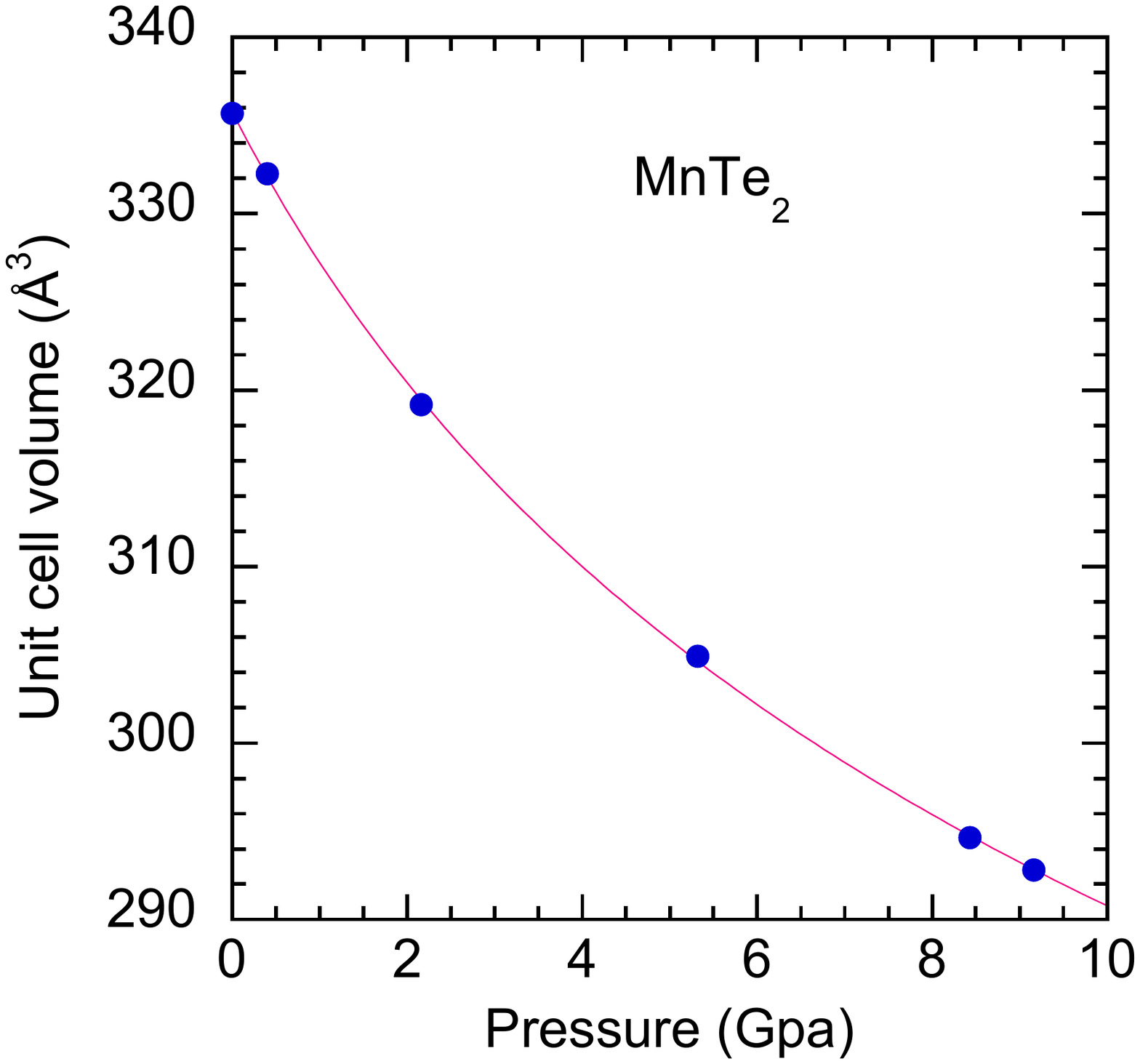}}
\vspace{5pt}
 \caption {(Color online) Pressure variation of the unit cell volume of MnTe$_2$ and its fit with Murnaghan equation of state that gives $B_0 = 34.6 \pm 1.0$ GPa and $Bo^{\prime} = 8.8 \pm 0.4$.  }
 \label{murnaghan}
\end{figure}

A compound of interest in this context is MnTe$_2$. MnTe$_2$ belongs to a large class of pyrite type and related marcasite and arsenopyrite type compounds MX$_2$ (M = transition element, X = Chalcogen or pnictogen element) with diverse magnetic and electrical properties. They range from insulator to metal or even superconductor. They can be diamagnetic, weakly paramagnetic, ferromagnetic or antiferromagnetic etc. The magnetic semiconductor MnTe$_2$ having a pyrite type crystal
structure, as shown in Fig. \ref{structure}, orders below $T_N \approx 88$ K in type-I antiferromagnetic
structure\cite{hastings59,chattopadhyay87a,burlet97} with the propagation vector ${\bf k} = (1,0,0)$. The magnetic phase
transition at $T_N$ was found to be of second order within experimental resolution \cite{chattopadhyay87a,burlet97} although
the related other manganese dichalcogenides MnS$_2$ and MnSe$_2$ undergo first-order phase transitions\cite{hastings76,chattopadhyay84,chattopadhyay91,chattopadhyay87b} at $T_N$. The
magnetic structure of MnTe$_2$ had been subject to controversy,\cite{hastings59,pasternak69,hastings70} regarding whether
the magnetic structure of MnTe$_2$ is of collinear single-{\bf k} or non-collinear triple-{\bf k} type, or whether there is
any spin reorientation transition. Burlet {\it et al.}\cite{burlet97} resolved this controversy and determined the
magnetic structure to be of non-collinear triple-{\bf k} type. The structure was found to be stable below $T_N$ down to 4.2 K,
the lowest temperature at which the magnetic structure was investigated. 

The high pressure X-ray diffraction was carried out previously,\cite{fjellvag85,fjellvag95} to study the pressure
induced volume changes in MnTe$_2$, though no detailed structural analysis was carried out in terms of determination of atomic
positions. Also, in a separate study the pressure dependence of Ne\'el temperature was obtained from resistivity and  
M\"ossbauer  measurements.\cite{vulliet01} The results of these two studies put together show a large violation of Bloch's rule, which 
however has not been stressed before. More importantly, a microscopic understanding of this phenomena was lacking.

In the present study, we take up this issue by experimentally revisiting the pressure dependence of the structure
and magnetic ordering temperature of MnTe$_2$ in terms of high-pressure powder neutron diffraction measurment, together with first-principles 
density functional theory (DFT) calculation to provide the microscopic understanding. The neutron diffraction study carried out in the 
present work, is undoubtedly a more reliable tool to measure the magnetic transition temperatures, compared to resistivity or
 M\"ossbauer. In addition, the present neutron diffraction 
study provide the detailed structural information, which was not available before, based on which our first-principles calculations
have been carried out. Our rigorous study confirms and rationalizes the breakdown of Bloch's rule in MnTe$_2$.

\section{Methodology}
High pressure neutron diffraction investigations were done on three neutron powder diffractometers, viz.  PEARL at the ISIS Facility in UK, D20 of Institute Laue-Langevin, Grenoble and also SNAP at SNS, Oak Ridge. Pressure was generated by Paris-Edinburgh pressure cells \cite{besson92,klotz05} and a mixture of 4:1 deuterated methanol:ethanol was used as pressure transmitting medium. The PEARL measurements used anvils made of tungsten carbide and a scattering geometry which restricted the available d-spacing range to below 4.2 $\AA$, i.e. the magnetic (100) reflection was not recorded. Rietveld refinements of the patterns to the crystal structure were carried out by the GSAS program \cite{vondreele86}. The experiments on D20 and SNAP used anvils made of cubic boron nitride \cite{klotz05} and a scattering geometry which gives access to reflections with larger d-spacings. The sample temperature was controlled using closed-cycle cryostats; fast cooling to 77 K was achieved by flooding the cell assembly with liquid N$_2$. The pressure was determined from the known pressure variation \cite{fjellvag85,fjellvag95} of the lattice parameter of MnTe$_2$.

DFT calculations on the experimentally measured structures were carried out in the plane-wave basis, within the
generalized gradient approximation (GGA) for the exchange-correlation functional, as implemented in the Vienna Ab-initio
Simulation Package.\cite{vasp} We used Perdew-Burke-Ernzerhof implementation of GGA.\cite{pbe} The projector augmented wave
potential was used. For the total energy calculation of different spin configurations, we considered a 2 $\times$ 2 $\times$ 1
super-cell containing a total of 48 atoms in the cell. For the self-consistent field calculation, an energy cut-off of 600 eV
and 4 $\times$ 4 $\times$ 8 Monkhorst-pack K-point mesh were found to provide good convergence of the total energy. The missing
correlation at the Mn sites beyond GGA, was taken into account through supplemented Hubbard $U$ (GGA+$U$)
calculation\cite{gga+u} following the Dudarev implementation, with choice of $U$ = 5.0 eV and Hund's coupling, J$_H$ of 0.8 eV.
Variation in $U$ value has been studied and found to have no significant effect on the trend.

\begin{figure}
 \resizebox{0.5\textwidth}{!}{\includegraphics{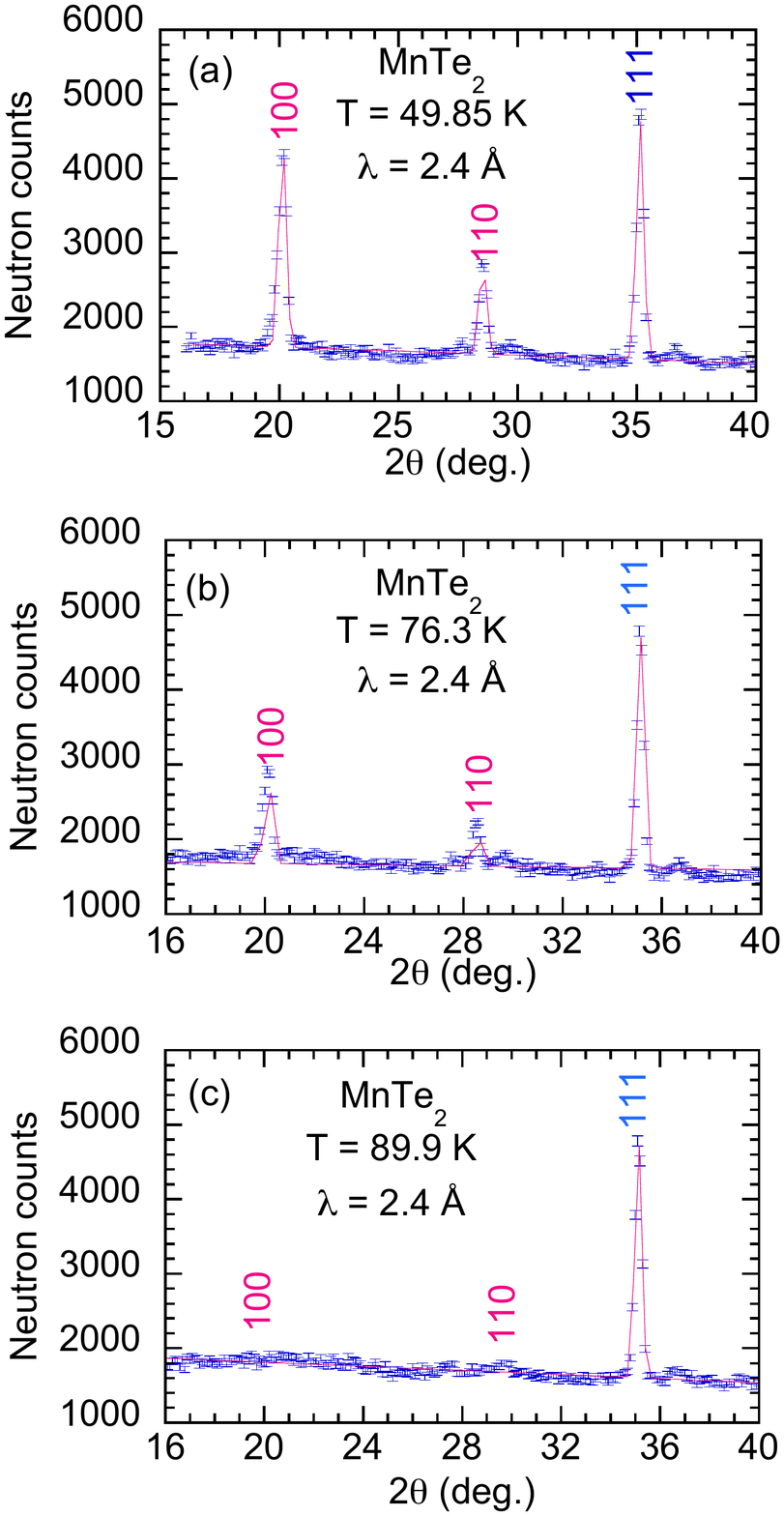}}
\vspace{5pt}
 \caption {(Color online) Neutron powder diffraction intensities of the $100$ and $110$ magnetic peaks along with the $111$ nuclear peak at several temperatures below and close to the antiferromagnetic N\'eel temperature $T_N = 88$ K. The data have been measured on the D20 diffractometer of the Institut Laue-Langevin in ambient pressure. The red lines are results of Gaussian fits of the Bragg peaks.}
 \label{ambient}
\end{figure}

\begin{figure}
\resizebox{0.5\textwidth}{!}{\includegraphics{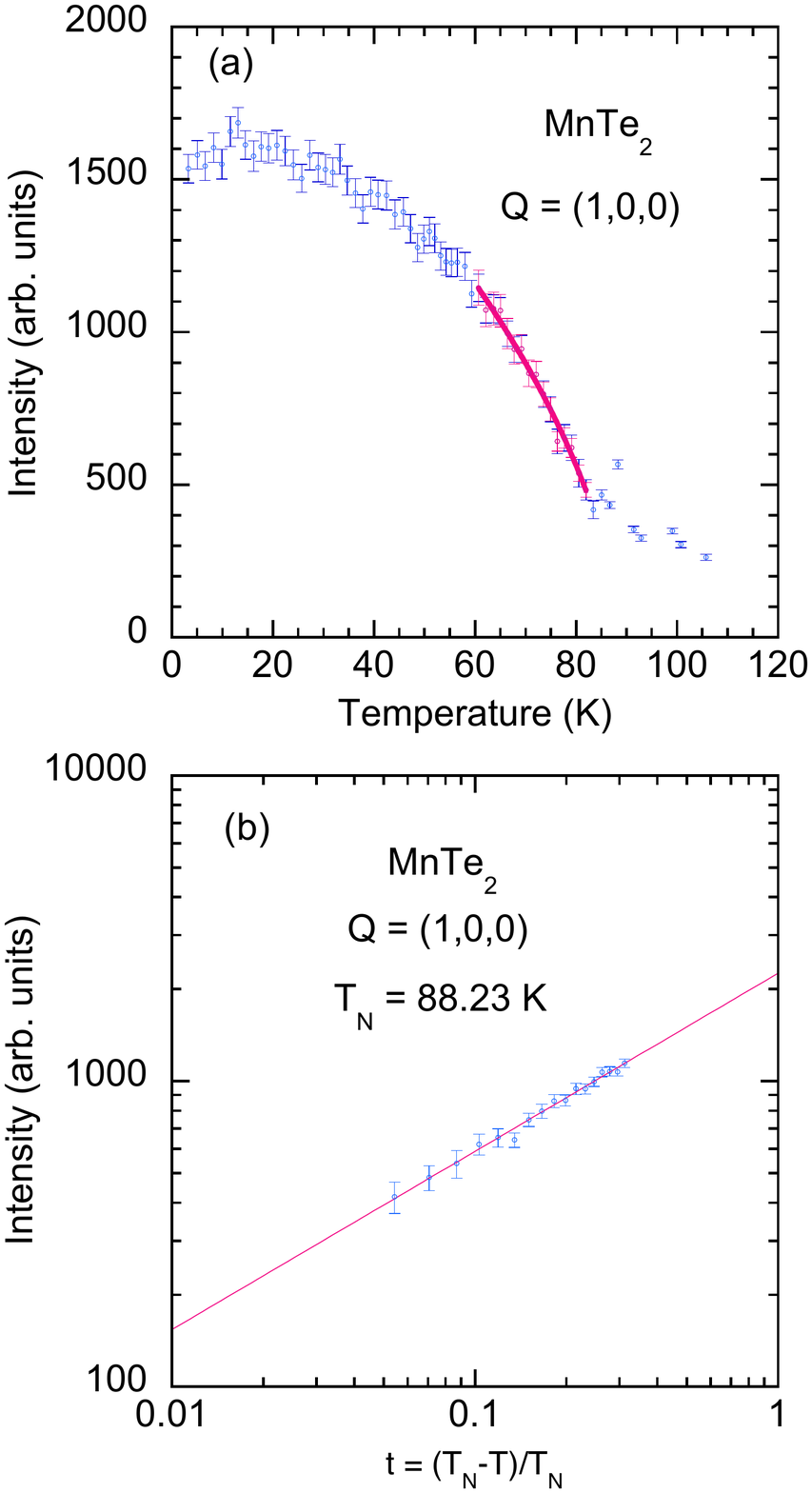}}
\vspace{5pt}
 \caption {(Color online) (a) Temperature variation of the integrated intensitiy of the $100$ magnetic Bragg peak. The red line is result of the least-squares fit of the data. The temperature range of the data used for the fit is also given by the temperature range of the red line. The least-squares fit gives the power-law exponent $ \beta = 0.29 \pm 0.04$
and a N\'eel temperature $T_N = 88 \pm 2$ K. (b) Log-log plot of the intensity of the $100$ magnetic reflection vs. reduced temperature. }
 \label{critical}
\end{figure}

\section{Evolution of the Structural parameters under Pressure}

The pyrite type crystal structure of MnTe$_2$ in the $Pa\bar{3}$ space group has Mn atom at $4(a)~(000)$ and Te atom at $8(c) ~(xxx)$ position. The cubic lattice parameter $a$ and the Te positional parameter $x$ were refined along with the isotropic atomic displacement parameters of Mn and Te atoms. Figure \ref{param} shows the pressure dependence of the structural parameters of MnTe$_2$, viz. lattice parameter, positional parameter $x$ of Te atom, Mn-Te and Te-Te bond lengths and the Mn-Te-Mn and Mn-Te-Te bond angles. The results are very remarkable and contrary to our naive expectation that the Te-Te bond distance would continuously decrease with pressure. Instead the bond distance seems to increase slightly at lower pressure but after reaching a maximum at P = 2 GPa it decreases and becomes somewhat flat at about P = 10 GPa. The two bond angles also show anomalous pressure dependence. This is expected since all the relevant bond distances and angles are derived from the single Te positional parameter $x$ and the cubic lattice parameter $a$ that decreases with pressure in the usual way. In contrast, the Mn-Te bond length is highly pressure sensitive and almost entirely responsible for the pressure-induced volume reduction, suggesting relatively weak Mn -Te bonds which
are susceptible to changes upon application of pressure.
The remarkable pressure response of the Te-Te bond distance and Mn-Te-Mn and Mn-Te-Te angles should be reflected in the pressure dependence of the superexchange interaction that decides the pressure variation of the N\'eel temperature, as obtained in our first-principles calculations. Fig. \ref{murnaghan} shows the pressure variation of the unit cell volume of MnTe$_2$ and its fit with Murnaghan equation of state. The fit gave $B_0 = 34.6 \pm 1.0$ GPa and $Bo^{\prime} = 8.8 \pm 0.4$ where $B_0$ is the bulk modulus and $Bo^\prime$ is its pressure derivative. The values agree well with the values determined previously from high pressure X-ray diffraction \cite{fjellvag85,fjellvag95}.

\begin{figure}
\resizebox{0.5\textwidth}{!}{\includegraphics{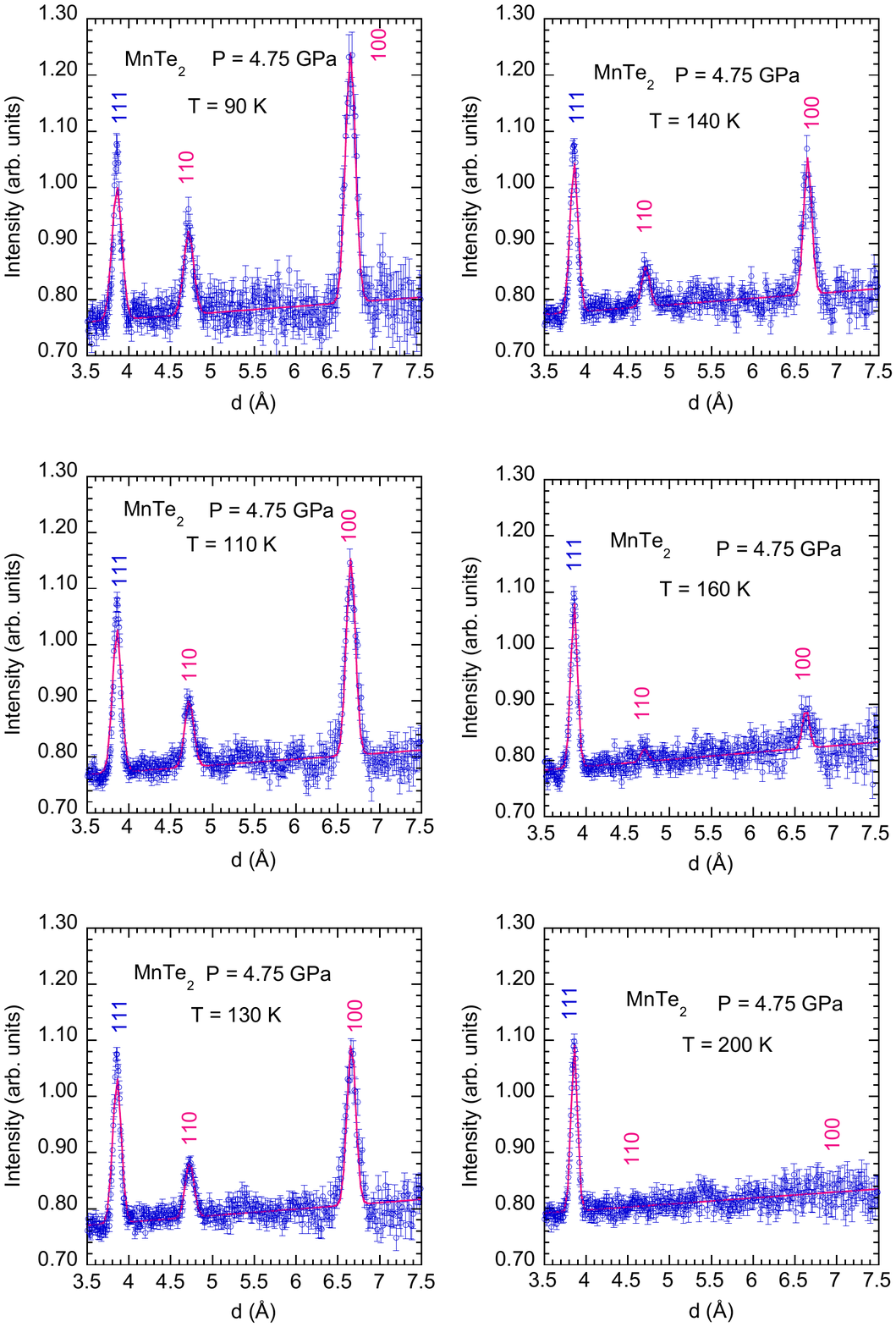}}
 \vspace{5pt}
 \caption {(Color online) Neutron powder  diffraction intensities of the $100$ and $110$ magnetic peaks along with the $111$ nuclear peak of MnTe$_2$ under $P = 4.75$ GPa at several temperatures below and above the antiferromagnetic N\'eel temperature measured on the SNAP diffractometer of the SNS of Oak Ridge National Laboratory. The red lines are results of Gaussian fits of the Bragg peaks. }
 \label{5GPa}
\end{figure}

 \begin{figure}
\resizebox{0.5\textwidth}{!}{\includegraphics{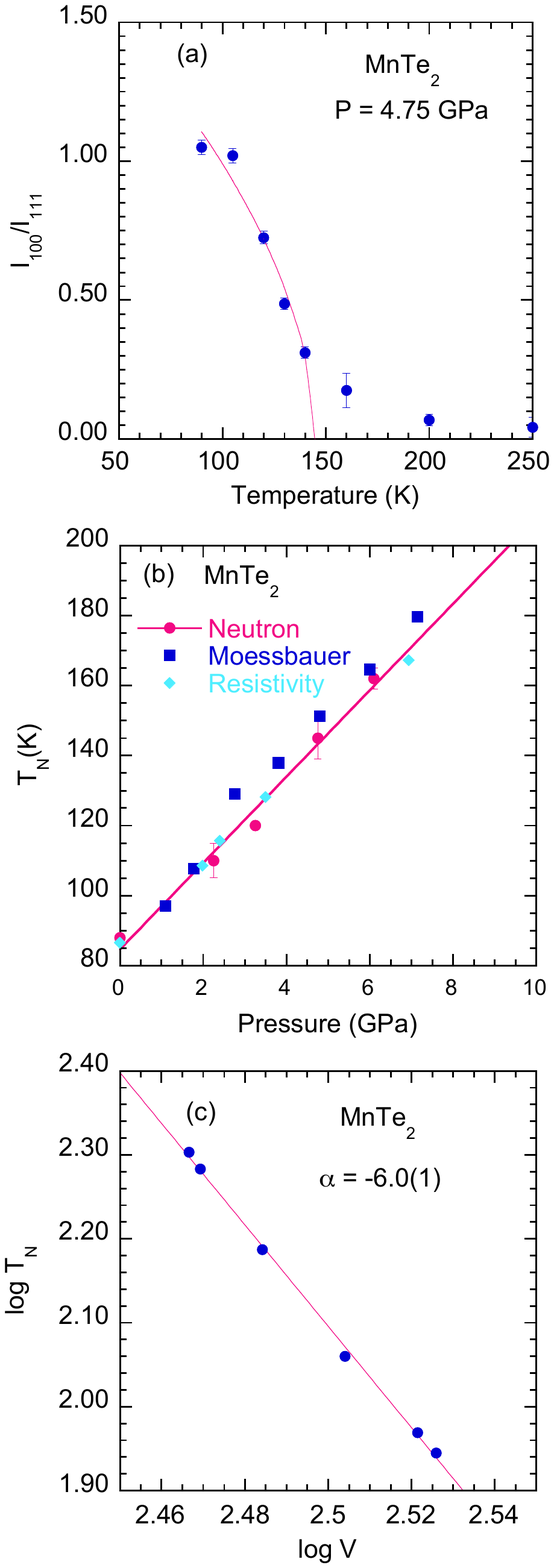}}
\vspace{5pt}
 \caption {(Color online) (a) Temperature variation of the intensity of the $100$ reflection of MnTe$_2$ at $P = 4.75$ GPa and its power-law fit. The red line is result of the least-squares fit of the data.  The power law fit yields $T_N =  145 \pm 7$ K. (b) Pressure variation of the N\'eel temperature of MnTe$_2$. The M\"ossbauer and the resistivity data are taken from Vulliet et al. \cite{vulliet01}.  The linear fit of the neutron data yields $\frac{dT_N}{dP} = 12$ K GPa$^{-1}$. (c) Log-log plot of N\'eel temperature T$_N$ vs. unit cell volume of MnTe$_2$. The red straight line is the result of the linear fit of the data yielding $\alpha =  -6.0 \pm 0.1$. }
 \label{bloch}
\end{figure}

\section{Measured Pressure Variation of T$_N$}

The antiferromagnetic phase transition of MnTe$_2$ was first investigated at ambient pressure with the sample (outside the pressure cell) fixed to the cold tip of the standard orange cryostat. Fig. \ref{ambient} shows neutron powder diffraction intensities of the $100$ and $110$ magnetic peaks along with the $111$ nuclear peak at several temperatures below and close to the antiferromagnetic N\'eel temperature $T_N = 88$ K. Fig. \ref{critical}(a) shows the temperature variation of the integrated intensity of the $100$ magnetic Bragg peak. The intensity of this reflection decreases continuously with
increasing temperature and becomes zero at about $T_N \approx 88$ K. The data just below $T_N$ could be fitted by a power-law exponent
\begin{equation}
I=I_0\left(\frac{T_N-T}{T_N}\right)^{2\beta}
\end{equation}
where $I$ is the integrated intensity, $I_0$ is the saturation value of the intensity at $T = 0$,
$T_N$ is the critical temperature and $\beta$ is the power-law exponent. Least-squares fit
of the data in the temperature range from $T = 60 $ to $T = 88$ K gave $ \beta = 0.29 \pm 0.04$
and a N\'eel temperature $T_N = 88 \pm 2$ K. The fitted value of the N\'eel temperature was used to determine the reduced temperature $t = (T_N-T)/T_N$. We then produced a standard log-log plot shown in Fig. \ref{critical} (b) to extract the critical exponent $ \beta= 0.29$ from the slope that agreed  well with that determined by the least-squares fit.

Fig. \ref{5GPa} shows neutron powder diffraction intensities of the $100$ and $110$ magnetic peaks along with the $111$ nuclear peak of MnTe$_2$ under $P = 4.75$ GPa at several temperatures below and above the antiferromagnetic N\'eel temperature. It is immediately noticed that the application of hydrostatic pressure P = 4.75 GPa increases the N\'eel temperature $T_N = 88$ K of MnTe$_2$ substantially. By fitting the temperature dependence of the intensity of the $100$ magnetic peak and fitting the data by a power law we determine $T_N =  145 \pm 7$ K. Fig. 7(a) shows this fit. Similarly we determined the N\'eel temperatures of MnTe$_2$ at several pressures. The result is shown in Fig. 7(b). The obtained trend agrees well with that obtained from resistivity and
  M\"ossbauer spectroscopy\cite{vulliet01}, as also shown in Fig. 7(b). The neutron diffraction results show that $T_N$ of
   MnTe$_2$ increases linearly in the pressure range $0-8$ GPa at a rate of about $12$ K GPa$^{-1}$,
    determined from the slope of the linear plot. From this linear relationship we calculated the T$_N$  values for the pressures at which
we determined the lattice and positional parameters of MnTe$_2$  from the high pressure neutron
     diffraction experiment on the PEARL diffractometer.  Fig. 7(c)  shows the log-log plot of N\'eel
      temperature T$_N$ vs. unit cell volume of MnTe$_2$. The slope of this plot gives $\alpha = -6.0 \pm
 0.1$ which is much larger than the Bloch rule value of $\alpha = -3.3$. Our result therefore point towards a
spectacular breakdown of Bloch's rule in MnTe$_2$. We note that transition to a non-magnetic state of the Mn$^{2+}$ ions
in MnTe$_2$ was reported\cite{vulliet01} from the resistivity and M\"ossbauer study,
and also evidenced by the pressure variation of infrared reflectivity investigated by Mita et al. \cite{mita08}.
Our experiments however did not show the volume collapse observed in high pressure X-ray diffraction
experiments \cite{fjellvag85,fjellvag95}. It is therefore plausible that we did not reach the transition pressure
during the present high pressure neutron diffraction experiments. The exact pressure at which
the transition to a non-magnetic state is expected to happen depends sensitively on the experimental conditions.

The present neutron diffraction data contain in principle the magnetic moment information because neutron diffraction probes both crystal and the magnetic structures and the intensities of the magnetic reflections when put to the absolute scale by using the intensities of the nuclear  reflections can give the ordered moment values. However this is not an easy task especially in a high pressure experiment using a large Paris-Edinburg pressure cell. The high absorption of the pressure cell and also a very high background hinders accurate determination of the nuclear and magnetic intensities. In the present case despite our efforts the determination of the pressure dependence of ordered moment from the neutron diffraction data was not successful.  We know however from our calculations (using pressure dependence of the structural parameters) that the ordered moment of Mn ions does not change at all (or very little) in the range 0 - 9 GPa investigated. The present high pressure neutron diffraction data seem to support this result. The intensity ratio of the magnetic and nuclear Bragg peaks do not change very much and is within the accuracy in the pressure range investigated.

\section{First-Principles Study}

\subsection{Calculated Pressure Variation of Ne\'{e}l Temperature}

\begin{figure}
\resizebox{0.45\textwidth}{!}{\includegraphics{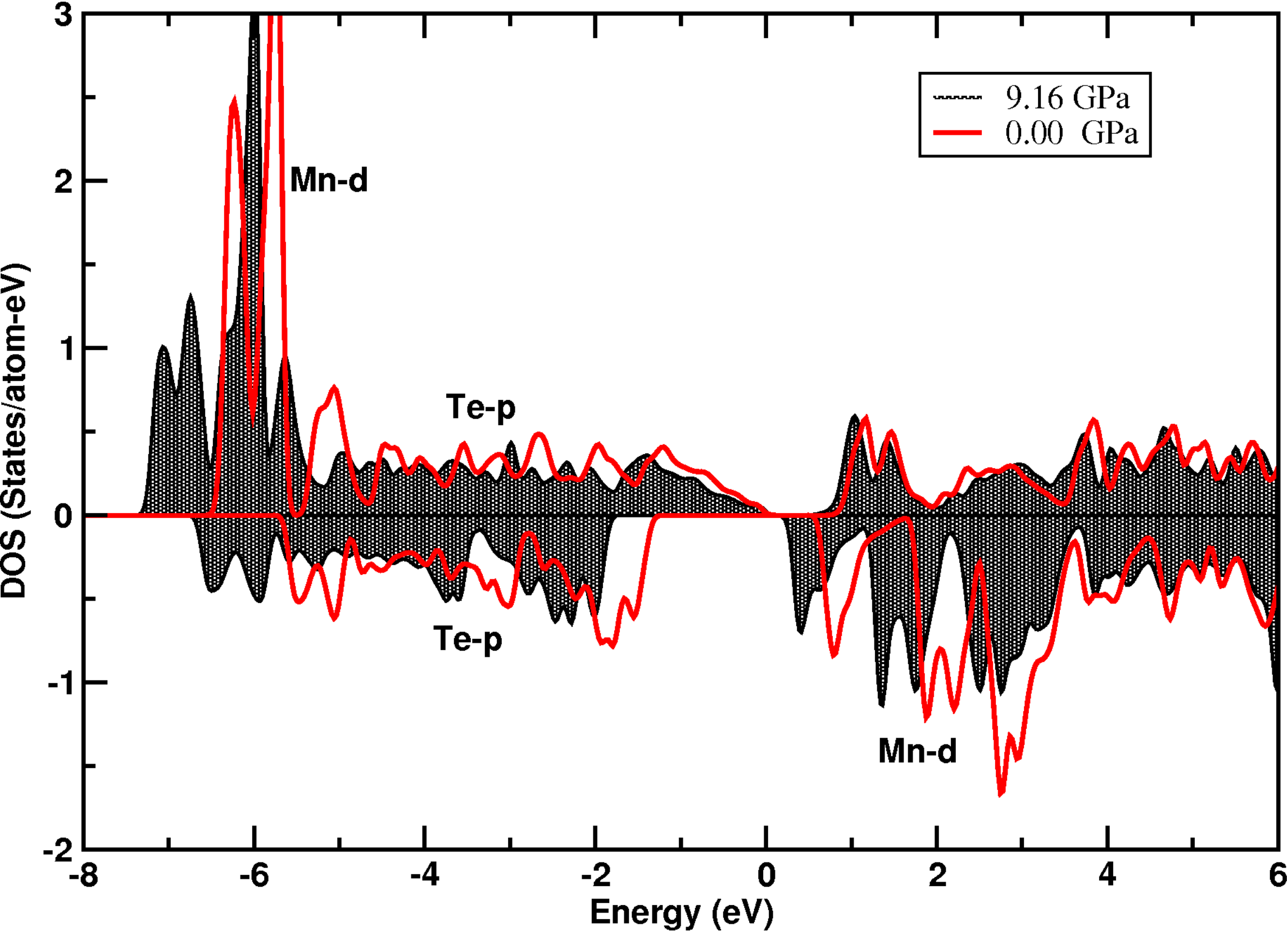}}
\label{bloch}
\vspace{5pt}
 \caption {(Color online) Comparison of GGA+$U$ density of states between the ambient pressure (solid line) and high pressure (shaded area)
phases. The energies are measured with respect to GGA+$U$ Fermi energy. The states of dominant Mn-$d$
and Te-$p$ characters have been marked.}
 \label{Fig2}
\end{figure}

To gain understanding on the significantly large pressure dependence of the Ne\'el temperature in 
MnTe$_2$ we carried out theoretical investigation in terms of first-principles DFT calculations. Fig. \ref{Fig2}
shows the comparison of the spin-polarized density of states of MnTe$_2$ at ambient pressure and at a pressure of
9.16 GPa, the highest pressure studied in the present calculations. The spin-polarized calculations within GGA+$U$
gave rise to a magnetic moment of 4.6 $\mu_B$ (4.5 $\mu_B$) at Mn site together with a moment of
0.03 $\mu_B$ (0.05 $\mu_B$) at Te site for the ambient (P = 9.16 GPa) pressure condition, suggesting the
high-spin state of Mn at both ambient and high pressure conditions, in agreement with experimental findings.
Both ambient pressure and high pressure phases were found to be insulating, with a gap at Fermi energy, marked
as zero in the figure. The Mn-$d$ states are fully occupied in the majority spin channel and completely
empty in minority spin channel, in correspondence with high spin state of Mn in its nominal 2+ valence state.
The comparison of the density of states between ambient pressure and at high pressure though, shows
enhancement of the Mn-$d$ band width by $\approx$ 1 eV, indicating the hopping interaction between
Mn-$d$ and Te$-p$ to increase substantially in moving from ambient to high pressure phase.

To extract the various magnetic interactions ($J$'s) between the Mn spins, we calculated the GGA+$U$ total energies
for various configurations of Mn spins and mapped the total energy onto an underlying S = 5/2 Heisenberg model.
Calculations were carried out for six different pressures, 0.0 GPa, 0.4 GPa, 2.16 GPa, 5.32 GPa, 8.43
GPa and 9.16 GPa. The dominant magnetic interactions considered in our calculation of $J$'s, were $J_1$, between the
first nearest neighbor (1NN) Mn atoms, connected to each other by the corner-shared Te atoms, and $J_2$, between the
second nearest neighbor (2NN) Mn atoms, connected to each other through Te-Mn-Te bridges.

\begin{figure}
\resizebox{0.45\textwidth}{!}{\includegraphics{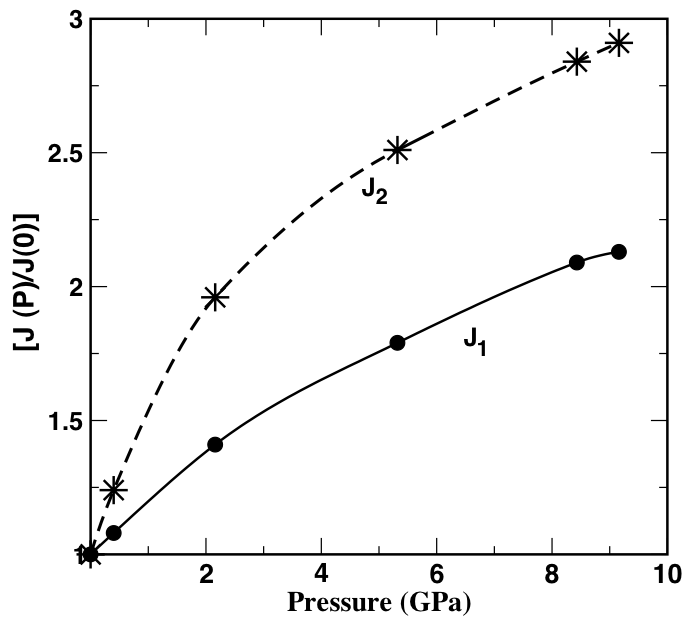}}
\label{bloch}
\vspace{5pt}
 \caption {The ratio of $J$'s at pressure P and that at ambient pressure (denoted as P = 0), plotted as a function of pressure.
The solid line corresponds to $J_{1}$, and dashed line corresponds to $J_{2}$.}
 \label{Fig4}
\end{figure}

Apart from the ferromagnetic (FM) configuration, with all Mn spins in the supercell pointing in the same direction, two
different antiferromagnetic (AFM) configurations, AFM1 and AMF2 were considered, with antiferromagnetic arrangement of
1NN Mn and 2NN Mn spins. The GGA+$U$ total energies corresponding to AFM configurations, measured with respect to the energy of FM configuration, turned
out to be negative, for all the studied pressures, in accordance with dominance of anti-ferromagnetic interactions.
Extracting J$_{1}$ and J$_{2}$ by mapping the total energy onto the spin Hamiltonian, given by
$H= -J{_1}\sum_{nn}S^{i}_{Mn}.S^{j}_{Mn} - J{_2}\sum_{2nn}S^{i}_{Mn}.S^{j}_{Mn}$, gave $J_{2}$ a small fraction of $J_1$,
with $J_{1}$/$J_{2}$ = 0.09 at ambient pressure and 0.12 at 9.16 GPa, suggesting the magnetism being primarily
governed by $J_1$. The pressure dependence of exchange interactions is shown in Figure \ref{Fig4}.

\begin{figure}
\resizebox{0.45\textwidth}{!}{\includegraphics{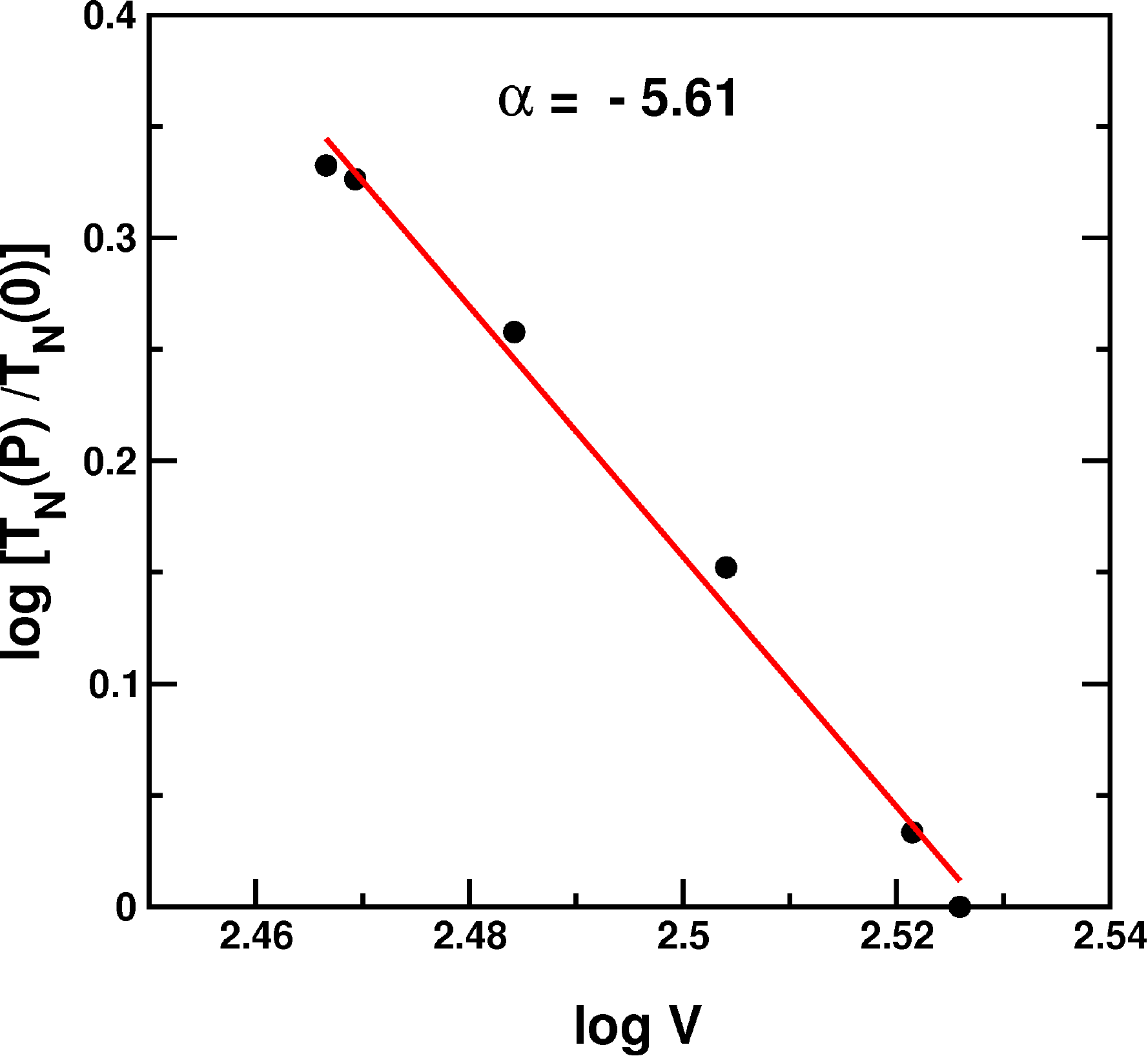}}
\label{bloch}
\vspace{5pt}
 \caption {(Color online) Log-log plot of T$_{N}$(P)/T$_{N}$(0) versus volume. The linear fit to the data points is shown with
line, having a slope of $\alpha$ = -5.61.}
 \label{Fig5}
\end{figure}

With the knowledge of J's, we calculated the Ne{\'e}l temperature T$_{N}$ using Mean-field
theory, given by $T^{mf}_{N}=\frac{S(S+1)J_{0}}{3K_{B}}$, where $J_{0}$ is the net effective interaction
12$J_{1}$ + 4$J_{2}$, S = 5/2, and $K_{B}$ is the Boltzmann constant. Mean field is expected to overestimate
the transition temperature, though the trend is expected to captured well, which is governed by $J$'s.
The computed log[$T_{N}$(P)/$T_{N}$(0)] plotted as a function of log($V$) is shown in Fig. \ref{Fig5}. The
straight line fit to the calculated data points gives rise to a slope of -5.61, close to the experimental
estimate of -6.0 $\pm$ 0.1. Both our experimental results and ab-initio calculations, thus establish
that Bloch's rule is largely violated in MnTe$_2$. In the following we theoretically investigate the microscopic
origin of this behavior.

\subsection{Microscopic Origin of breakdown of Bloch rule}

Bloch's rule has been found to be very successful with many magnetic insulators, especially
the transition metal oxides and fluorides. The question then arises: what makes
the Bloch's rule fail for MnTe$_2$? In order to explore this, we extracted
the hopping interactions $b^{ca}$, where $c$ and $a$ signifies Mn and Te, respectively,
by carrying out N-th order muffin-tin orbital (NMTO) based NMTO-downfolding calculations.\cite{nmto}
This involved construction of real-space representation of the Mn $d$- O $p$ Hamiltonian
in Wannier function basis out of the full DFT calculations, by integrating out all the degrees
of freedom other than Mn $d$ and Te $p$. Our NMTO-downfolding calculations
to extract the dependence of $b^{ca}$ on cation-anion distance $r$ gave,
$b^{ca} \sim \frac{1}{r^{4.2}}$, instead of $b^{ca} \sim \frac{1}{r^{2.5}}$
assumed for derivation of the Bloch's rule. This gives $T_N ~\sim \frac{1}{r^{17}} \sim \frac{1}{V^{5.67}}$,
very close to the estimate obtained from total energy calculations, as well as from the experiment.
This points to the fact that violation of Bloch's rule in case of MnTe$_2$ is caused due to the
deviation in the distance dependence of $b^{ca}$, from the $\frac{1}{r^{2.5}}$ behavior
rather than that by $U$ or $\Delta$. We find that the distance dependence of $b^{ca}$ found
for MnTe$_2$, is more like the canonical behaviour,\cite{oka} in which the interatomic matrix elements
are supposed to scale with distance as $\frac{1}{r^{l+l'+1}}$ where $l$ and $l'$  are the angular momenta
of the orbitals involved. 
In case of several TM oxides, on the other hand, anaylysis of DFT band structure,\cite{dd} gave rise to
a  $\frac{1}{r^{l+l'}}$ behavior, similar to that obtained from molecular orbital theory or configuration
interaction method on KNiF$_4$ or MnO or MnF$_2$\cite{smith69,shrivastava76}. This presumably originates
from differential nature of Te 4$p$ wavefunctions compared with that of 2$p$ or 3$p$, which together with 
non monotonic pressure dependence of position parameter, x, influences the super-exchange interaction in
a differential manner.

\section{Summary}

In conclusion, the N\'eel temperature of MnTe$_2$ was found to show unusually large
pressure dependence of about $12$ K GPa$^{-1}$, which has been confirmed in the present study   
through more rigorous and reliable high pressure neutron diffraction experiments compared
to that in literature, as well as through first-principles density functional theory calculations.
Our measured pressure dependence of the N\'eel temperature and unit cell volume
gave $\alpha = -6.0 \pm 0.1$ which is much larger than that expected from the Bloch's rule
$\alpha=\frac{d\log T_N}{d\log V}=-\frac{10}{3} \approx -3.3$. The calculated pressure dependence of N\'eel temperature
gave rise to $\alpha= -5.61$ in good agreement with the experimental estimate. We provided a microscopic
understanding of this behavior in terms of the distance dependence of Mn-Te hopping interaction
upon application of pressure, which showed significant deviation from that for NiF$_4$ or MnO or 
MnF$_2$\cite{smith69,shrivastava76}.

Finally, the large pressure dependence of magnetic interactions and magnetic ordering temperature 
provide us with a handle to tune the properties of magnetic materials, which can lead to important technological applications.
The present study, should have important bearing on this topic.

\end{document}